\documentclass[a4paper]{JAC2003}
\usepackage{graphicx}
\usepackage{booktabs}
\usepackage{subfig}
\usepackage{amsmath}
\usepackage{amsfonts}
\usepackage{url}
\usepackage{cite}

\usepackage{varwidth}
\usepackage{xcolor}

\begin{document}
\title{STABILITY DIAGRAM OF COLLIDING BEAMS}

\author{X. Buffat, EPFL, Lausanne; CERN, Geneva, Switzerland\\
W. Herr, N. Mounet, T. Pieloni, CERN, Geneva, Switzerland}

\maketitle

\begin{abstract}
The effect of the beam--beam interactions on the stability of impedance mode is discussed. The detuning is evaluated by the means of single particle tracking in arbitrarily complex collision configurations, including lattice non-linearities, and used to numerically evaluate the dispersion integral. This approach also allows the effect of non-Gaussian distributions to be considered. Distributions modified by the action of external noise are discussed.
\end{abstract}

\section{Introduction}
The stability of impedance driven modes is usually ensured by non-linearities via Landau damping. The strength of the Landau damping is represented in a stability diagram, defining the area of complex tune shifts which can be kept stable~\cite{general-sdiag}. The computation of the stability diagram is a crucial element for the evaluation of the non-linearity required for the beam stability. It can be done analytically in many simple cases, in particular considering lattice non-linearities and a limited number of beam--beam interactions. Facing the complexity of the LHC collision scheme, a numerical approach seems more appropriate. In particular, the dispersion integral is solved numerically using the amplitude detuning provided by single particle tracking, thus allowing the computation of stability diagrams in arbitrarily complex cases.

Following the same approach, the effect of non-Gaussian distribution functions can be introduced in the computation. Previous studies have shown that the distribution function plays a crucial role in the stability diagram~\cite{non-gaussian-sdiag} and, being usually poorly known, greatly diminishes the predicting power of such a consideration. The numerical evaluation of the stability diagram allows us to go further in the understanding of this effect by using a non-analytical distribution function. In particular, coloured external noise is known to enhance diffusion of resonant particles~\cite{resonant-diffusion}, greatly modifying the distribution and therefore the stability diagram.

\section{Stability of multibunch coherent beam--beam modes}
The method to derive the stability of impedance driven multibunch modes by placing their tune shifts in the stability diagram derived using the lattice non-linearities, as described in~\cite{multibunch-mode}, considers bunches with identical detuning. In the LHC, this assumption is no longer valid once beam--beam effects are considered. Indeed, the asymmetric layout of the interaction points as well as the asymmetric filling scheme lead to a variety of bunches having significantly different detuning. Theoretical treatment of the beam stability in such configurations is currently lacking. It is however possible to assess these cases using multiparticle tracking simulation~\cite{BBZ}, at the expense of large computational power.

\begin{figure}
 \centering
\subfloat[BBQ]{
\includegraphics[width=0.8\linewidth]{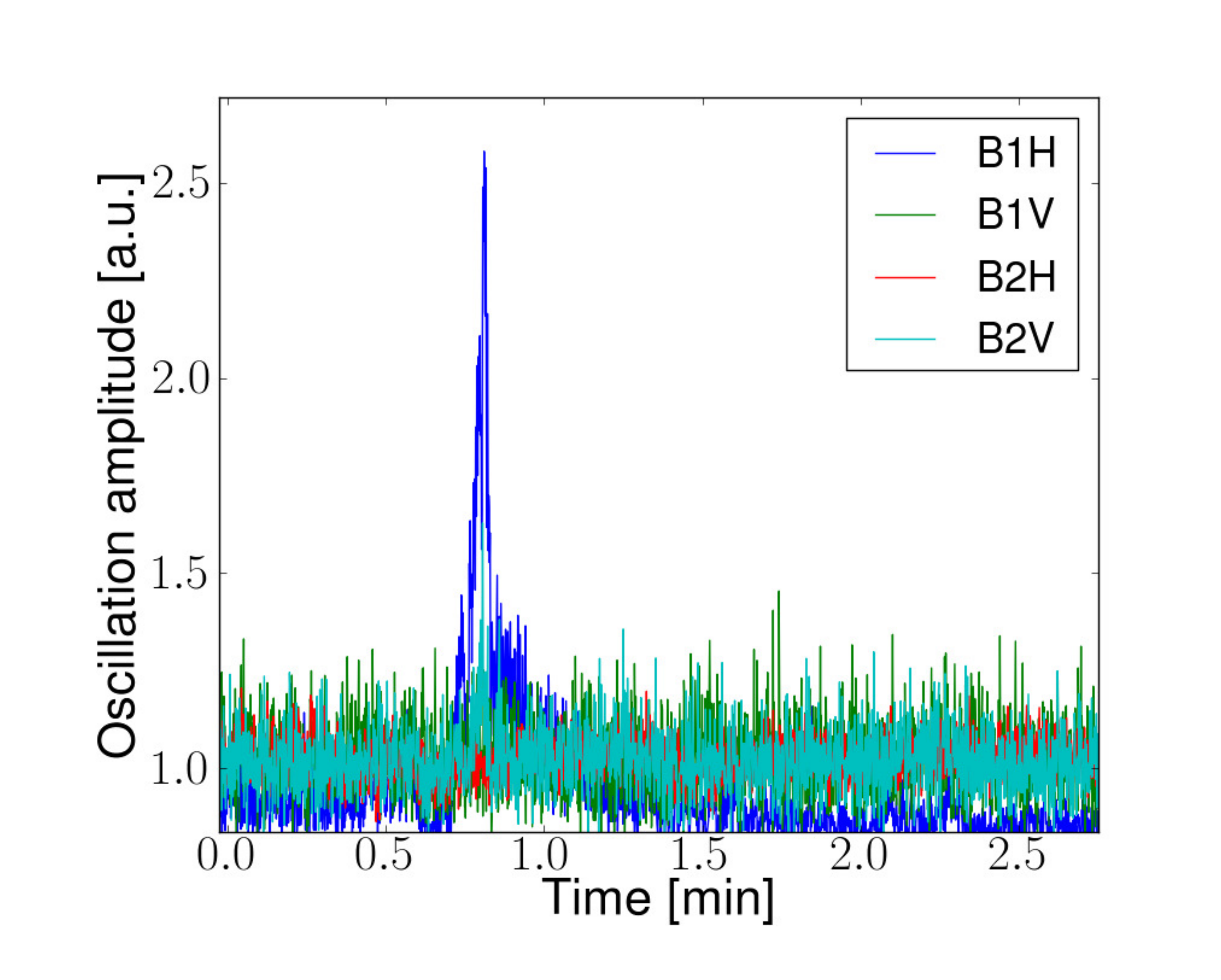}
}
\qquad
\subfloat[FBCT]{
\includegraphics[width=0.8\linewidth]{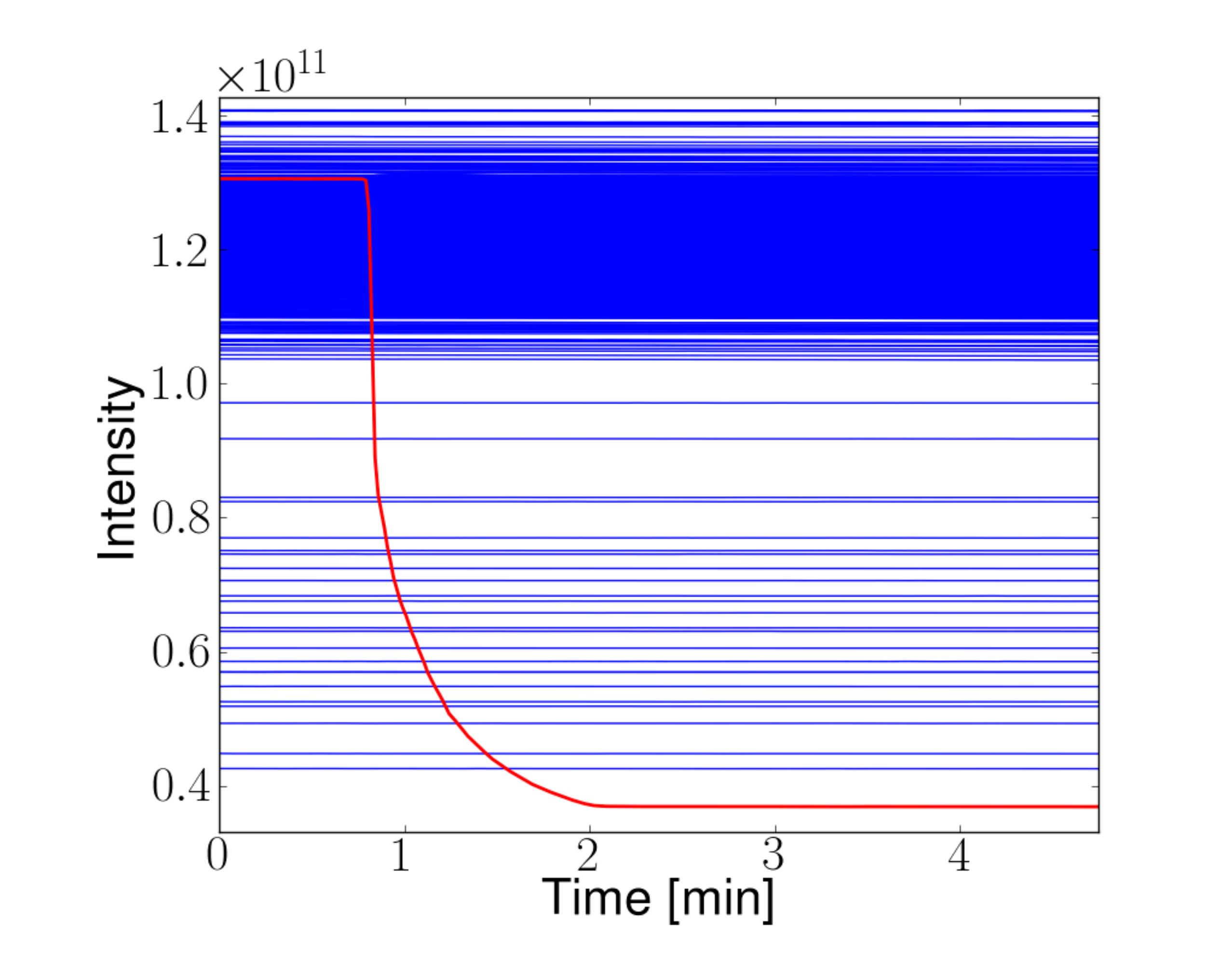}
}
\caption{Observation of an instability during luminosity production in the LHC (fill 2644). The beam oscillation amplitude shows a coherent excitation in the horizontal plane of Beam 1. The bunch by bunch intensity of both beams shows that all bunches (in blue) except one (in red) are stable.}
\label{fig-observation}
\end{figure}

The beam--beam interactions do not only introduce bunch dependent detuning, they also change the nature of the coherent modes. The Landau damping of beam--beam modes in the single bunch regime is addressed in~\cite{stability-BB-mode}, the extension to multibunch coherent beam--beam mode is, however, not trivial. Preliminary studies by the means of multiparticle tracking simulation are presented in~\cite{BBZ}. Such an approach is well suited to studying the LHC in its full complexity, however it is very demanding in terms of computing power. It is therefore interesting to consider simplified cases. In this paper, we address the stability of single bunch impedance modes with detuning from the lattice and beam--beam interactions, in other words, multibunch effects from the impedance as well as beam--beam coherent mode are neglected. These drastic assumptions are motivated by the observation, in the LHC, of single bunch instabilities, while operating in the multibunch regime. Figure~\ref{fig-observation} shows such an instability at the end of the squeeze during operation of the LHC in 2012. In particular, the measurement of the beam oscillation amplitude provided by the BBQ system indicates a coherent instability on Beam 1. The observation of bunch intensities indicates that only one bunch lost its intensity in an abnormal way, with respect to the other bunches, suggesting that only this bunch had undergone the instability.

\section{Numerical evaluation of stability diagrams}
Considering a case without coupling, the stability diagram of each plane is obtained by solving the dispersion relation for a given detuning $q(J_x,J_y)$ and distribution function $\psi(J_x,J_y)$ where $J_x$ and $J_y$ are the unperturbed actions in each plane,
\begin{equation}
 \frac{-1}{\Delta Q_i} = \iint_{0}^\infty \frac{J_i\frac{\mathrm{d}\psi}{\mathrm{d}J_i}}{Q-q(J_x,J_y)}\mathrm{d}J_x\mathrm{d}J_y,~Q\in \mathbb{R},~i = 1,2.
\end{equation}
 The $\Delta Q_i$ found for different values of $Q$ are the tune shifts at the limit of stability, therefore, they define an area in which the tune shifts are stable. In cases where the denominator is strictly non-zero, the tune shift is purely real, which indicates the absence of Landau damping. In the relevant cases, the integral possesses poles, and can be evaluated using various techniques. Whereas it is difficult to obtain analytically in complex configurations, the detuning is easily obtainable numerically. The dispersion integral can then be evaluated by standard numerical techniques, in our case by adding a vanishing complex part to the denominator.
\subsection{Detuning}
\begin{figure}
 \centering
\includegraphics[width=0.9\linewidth]{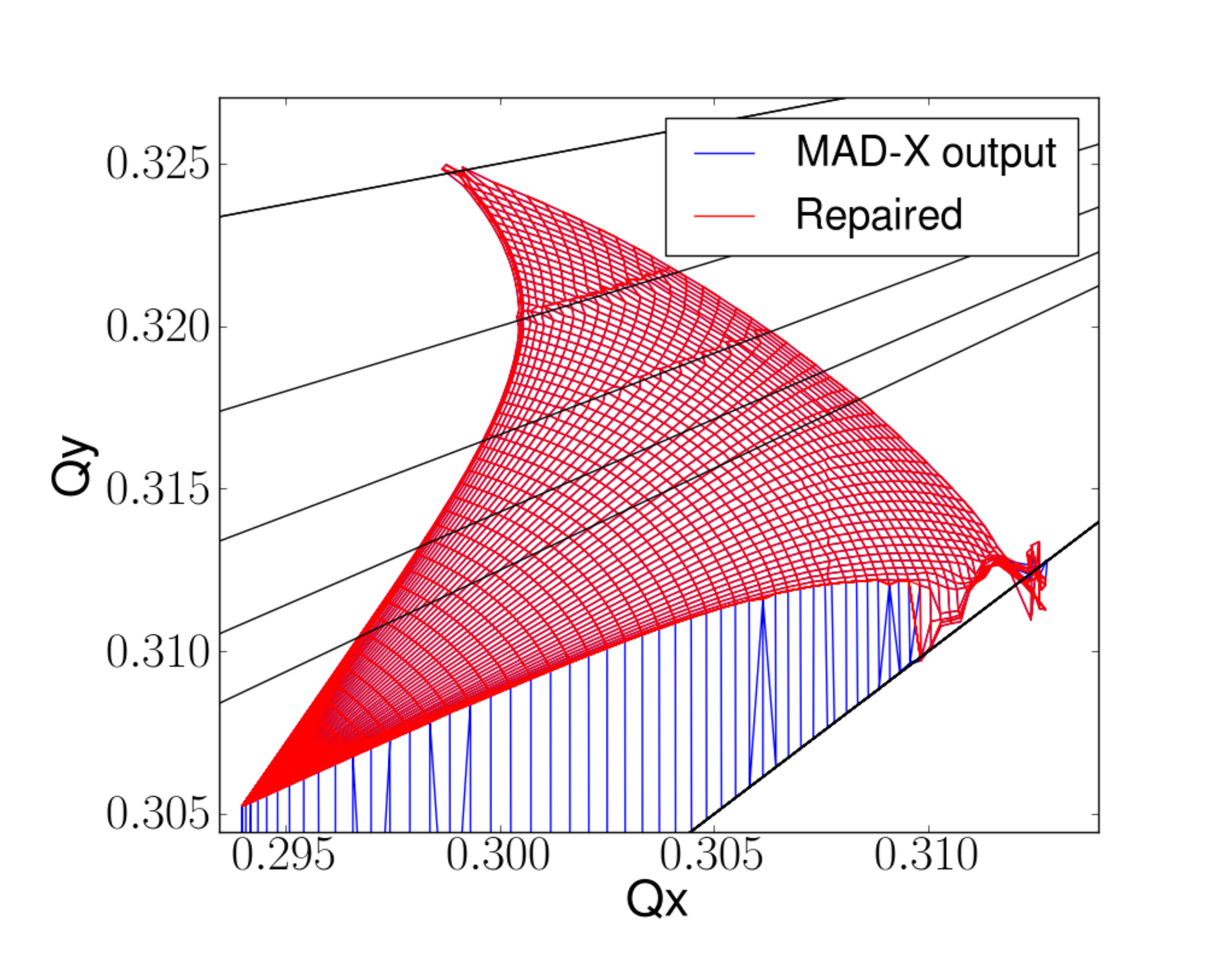}
\caption{Illustration of a typical tune footprint with octupoles (positive polarity) and long-range beam--beam interactions at the end of the squeeze in the LHC. The tracking result is often distorted by the presence of resonances; using a fine mesh and a repairing algorithm allows us to get a footprint suitable for the computation of stability diagrams.}
\label{fig-footprint}
\end{figure}
The detuning is obtained numerically from tracking simulation with MAD-X~\cite{madx,madx-footprint,madx-footprint2}. Particles with different actions are tracked for 1024 turns from which the tunes can be evaluated, the result is represented in the tune diagram as a tune footprint (Fig.~\ref{fig-footprint}). A significant number of particles is required to reduce the noise in the evaluation of the integrals. In particular, the effect of resonant driving term on the tracking can have a significant effect on the evaluation of the footprint. This effect is reduced by an automatic removal of faulty points and interpolation from the closest non-faulty points. As illustrated in Fig.~\ref{fig-footprint}, this technique does not allow us to fully remove distortions due to strong non-linearities, in particular from coupling resonances. In the cases of practical interests, however, the effect on the footprint, and consequently on the stability diagram, is tolerable. The integrity of the footprint is nevertheless systematically checked before deriving a stability diagram.
\subsection{Benchmarking}
\begin{figure}
 \centering
\includegraphics[width=0.9\linewidth]{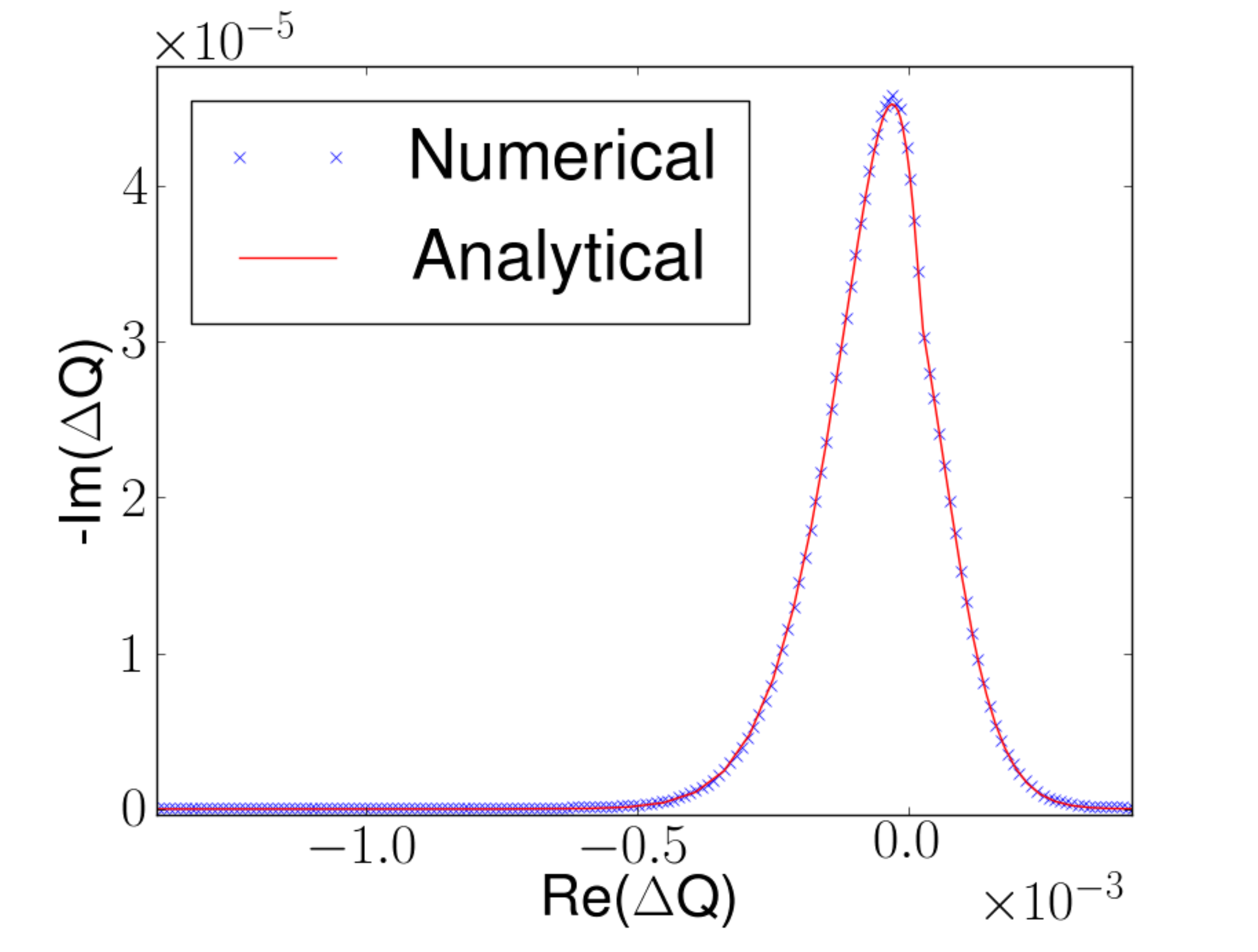}
\caption{Comparison of analytical and numerical derivation of a stability diagram with LHC octupoles powered with -100 A for a 4 TeV beam with a normalized emittance of 2 $\mu$m.}
\label{fig-benchmarking}
\end{figure}

Figure~\ref{fig-benchmarking} shows a good agreement between stability diagrams computed analytically and numerically in the case of octupolar detuning using a Gaussian distribution.

Practical applications of this code for the LHC can be found in~\cite{LHC-cases}.

\section{The effect of external noise}
\begin{figure}
 \centering
\subfloat[Relative difference to initial distribution in action space after $2\cdot10^4$ turns]{
\includegraphics[width=0.75\linewidth]{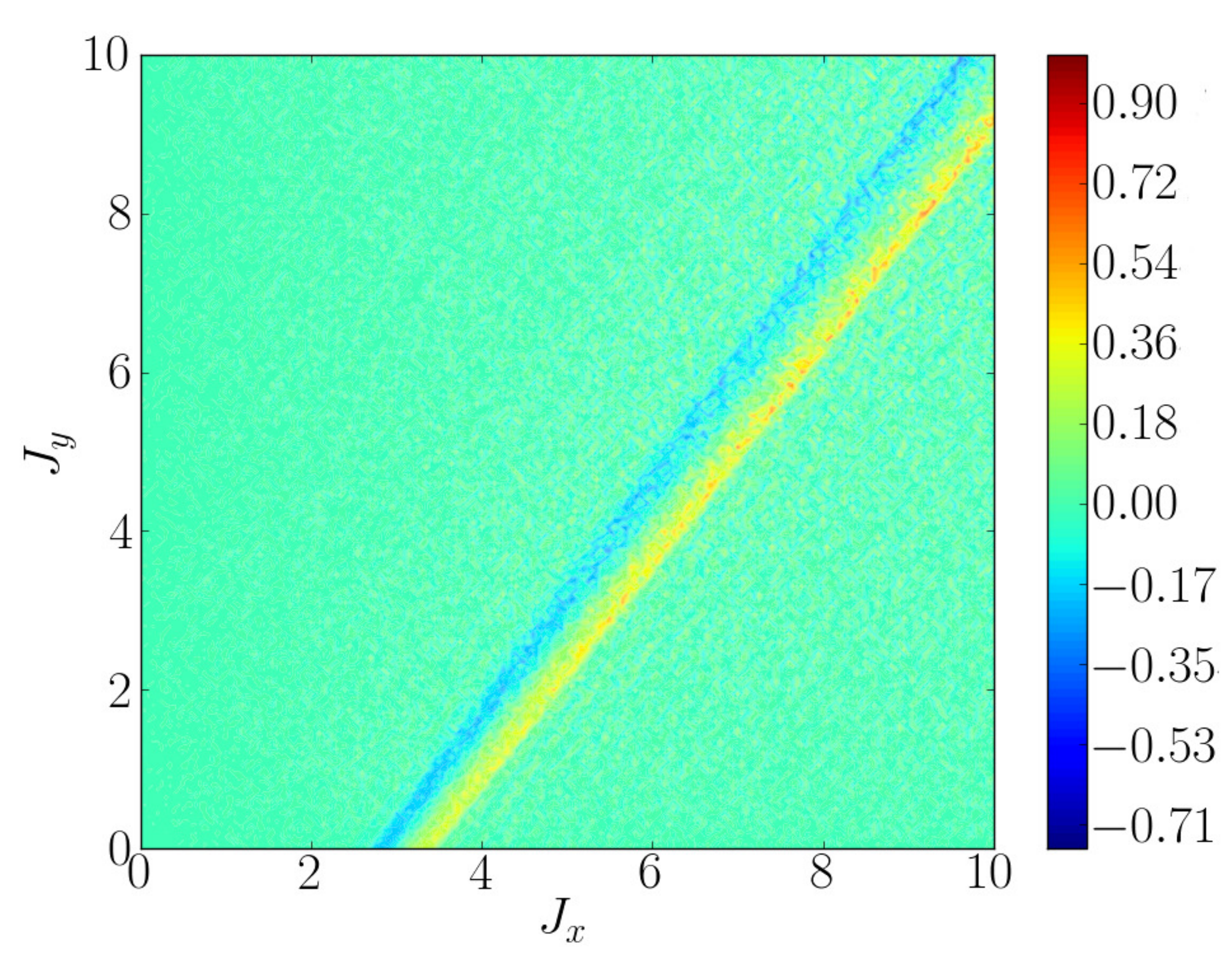}
\label{fig-distr noise}
}
\qquad
\subfloat[(Un)perturbed stability diagrams]{
\includegraphics[width=0.75\linewidth]{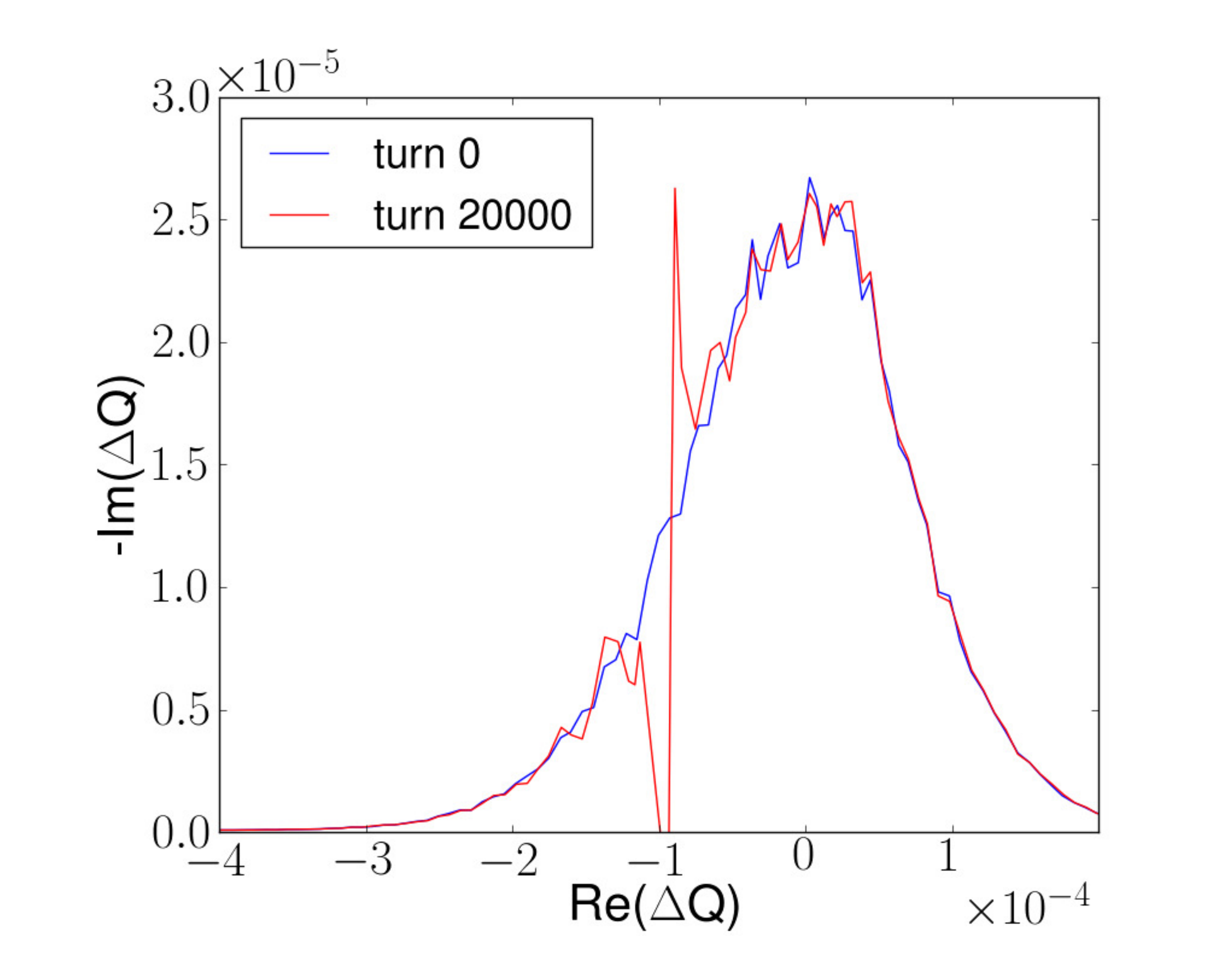}
\label{fig-sdiag noise}
}
\qquad
\subfloat[Real space bunch profile]{
\includegraphics[width=0.75\linewidth]{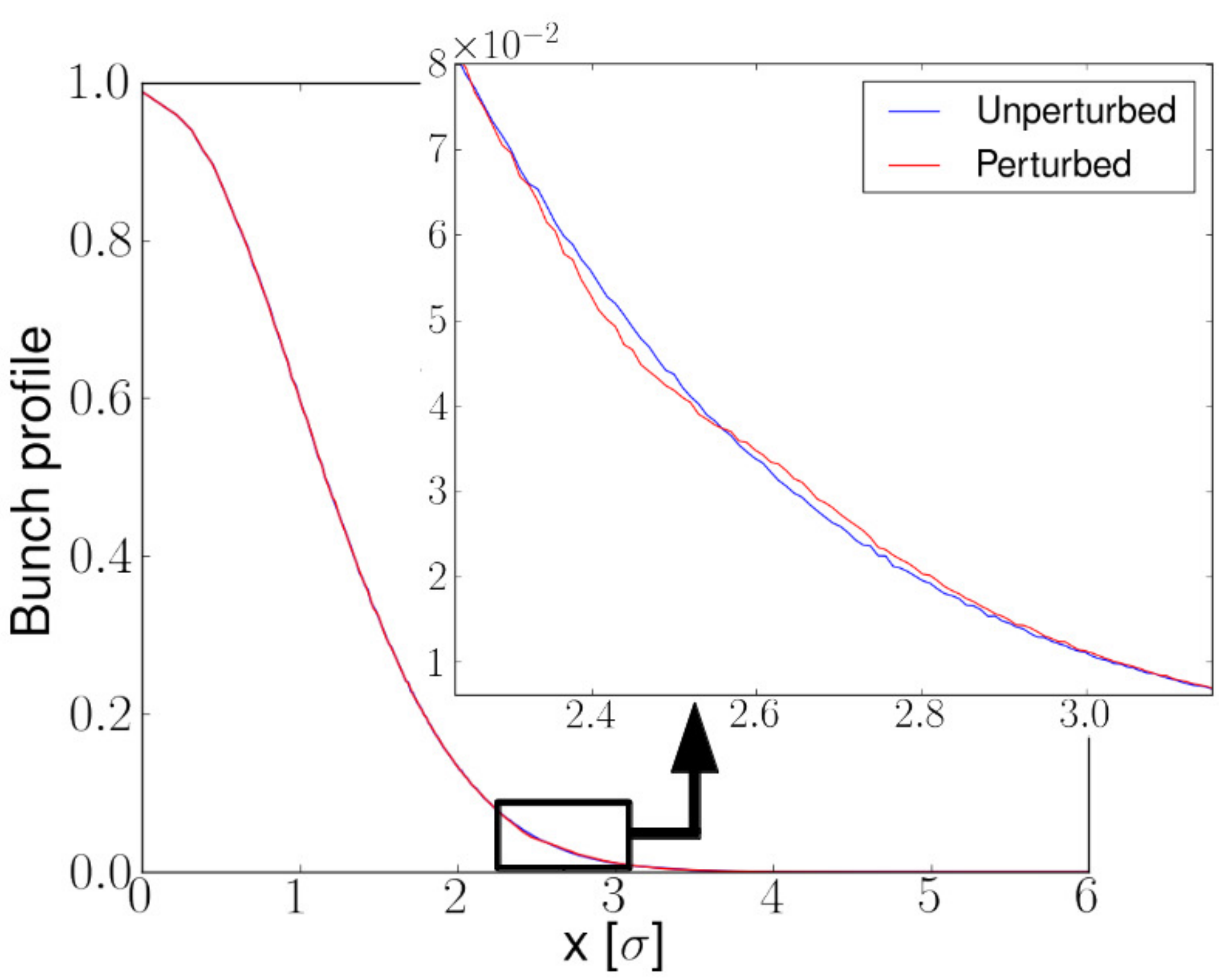}
\label{fig-distr real}
}
\caption{Stability diagram derived from a distribution perturbed by external noise in the presence of amplitude detuning $a=1.64\cdot10^{-4}, b=-1.16\cdot10^{-4}$. The noise has an amplitude of $10^{-4}\sigma_{x'}$ and a frequency of $Q_{x,0}+0.05$ with a correlation time of 100 turns.}
\label{fig-distr}
\end{figure}

External coloured noise enhances the diffusion of resonant particles. In the presence of amplitude detuning, this results in a depletion of certain parts of the distribution. This effect is illustrated with a simple model, using a multiparticle code that tracks particles through a lattice with amplitude detuning described by
\begin{eqnarray}
 Q_x &=& Q_{x,0} + a\cdot J_x  + b \cdot J_y, \\
 Q_y &=& Q_{y,0} + b\cdot J_x +  a \cdot J_y.
\end{eqnarray}
The noise is modelled with a sinusoidal excitation with finite correlation time. The relative difference to the initial distribution in action space after $2\cdot10^4$ turns is shown in Fig.~\ref{fig-distr}\subref{fig-distr noise}. The depletion of the distribution at the position of resonant particles is visible. This effect was already studied for different purposes in~\cite{distribution}. The measurable effect on the distribution in real space is very small (Fig.~\ref{fig-distr}\subref{fig-distr real}), the effect on the stability diagram is, however, significant, as illustrated by the comparison of the stability diagrams obtained using the initial and perturbed distributions (Fig.~\ref{fig-distr}\subref{fig-sdiag noise}).

\begin{figure}
 \centering
\subfloat[Stability diagram and tune shifts] {
\includegraphics[width=0.9\linewidth]{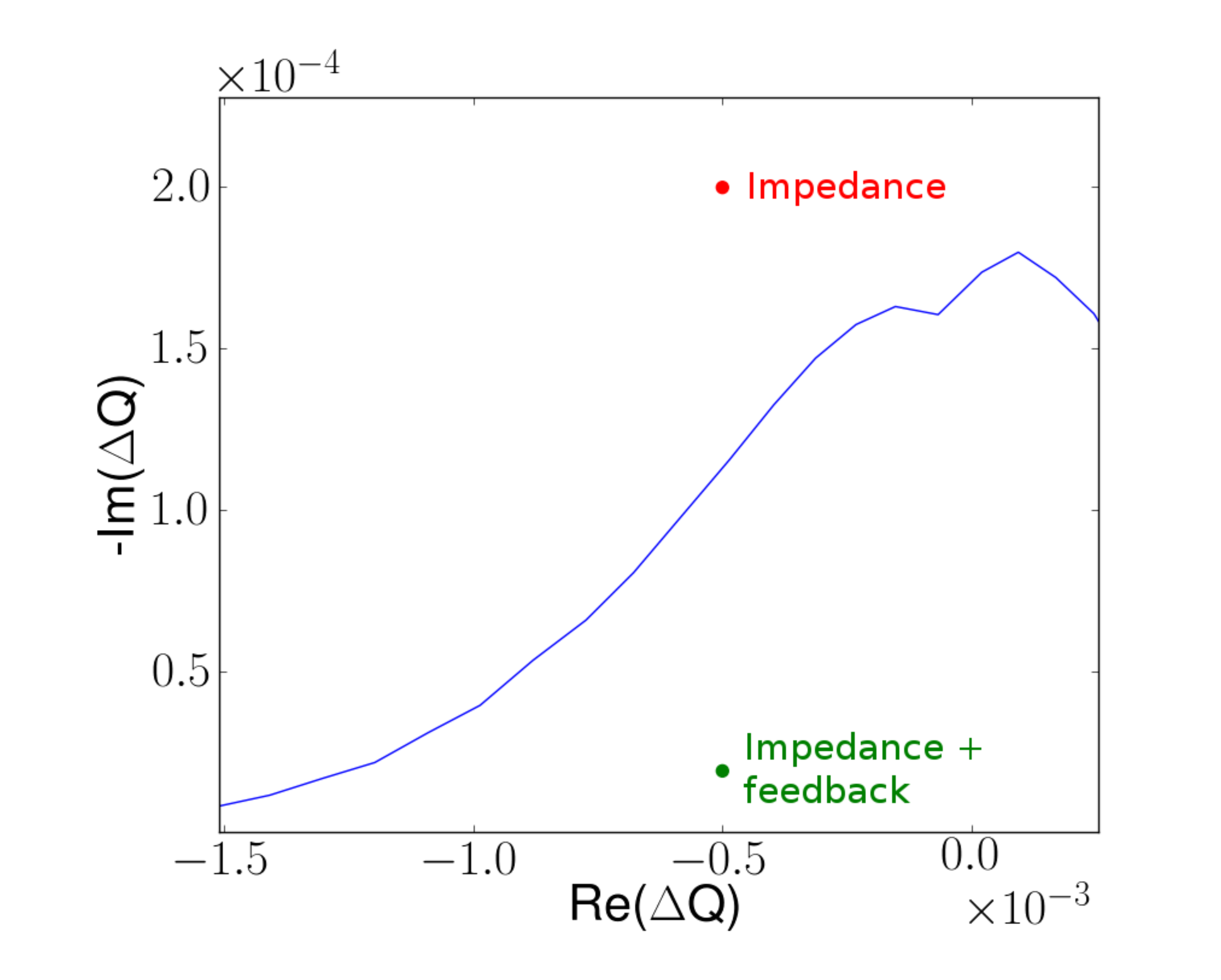}
\label{fig-study case sdiag}
}
\qquad
\subfloat[Tracking without external noise] {
\includegraphics[width=0.8\linewidth]{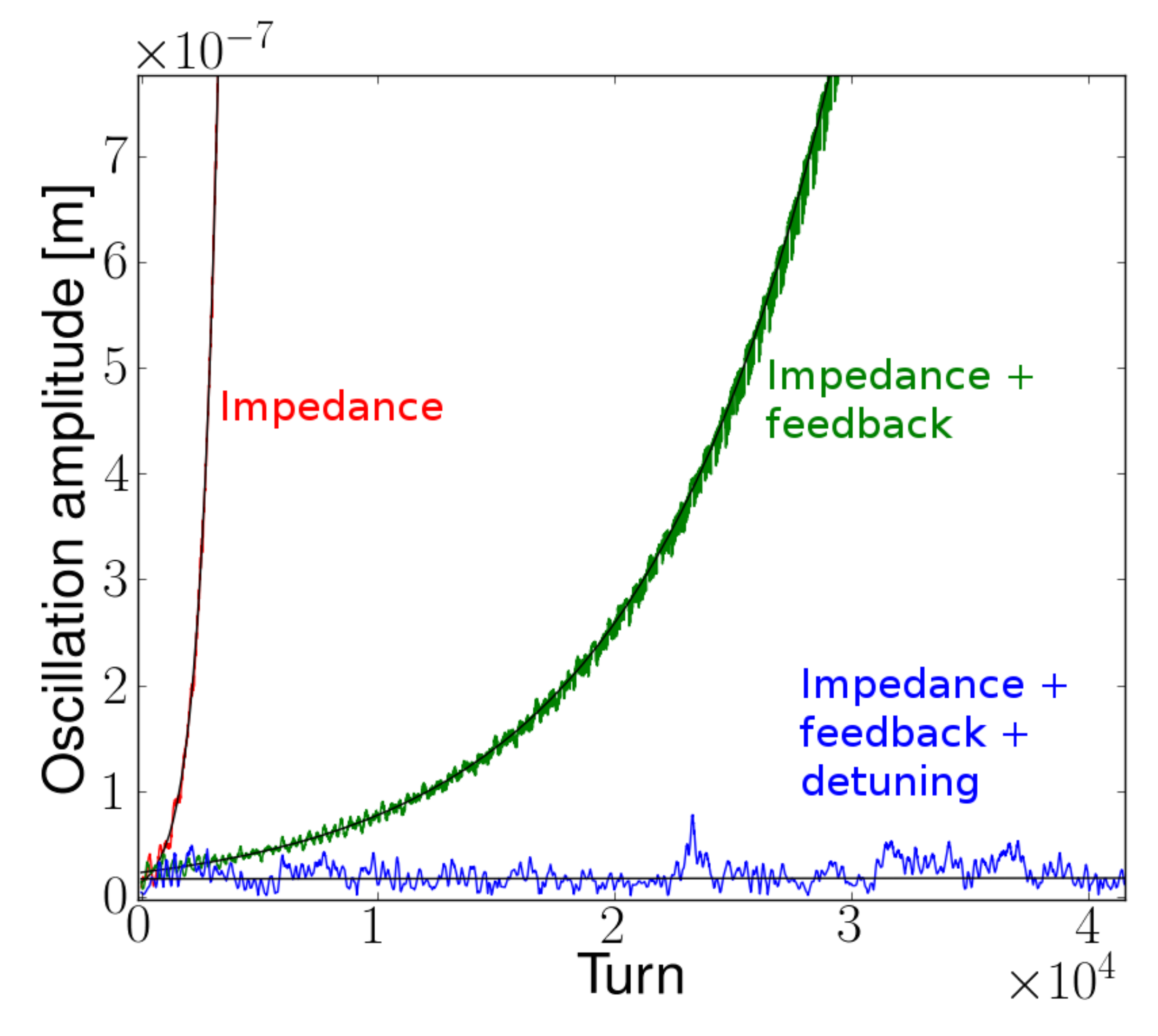}
\label{fig-study case rise}
}
\caption{Analysis of a single bunch with intensity $1.5\cdot10^{11}$ in the LHC, chromaticity of $-10$ units, stabilized with a transverse damper with gain $\sim700$ turns and amplitude detuning $a=-2.05\cdot 10^{-4},b=1.45\cdot10^{-4}$.}
\label{fig-study case}
\end{figure}

\begin{figure}
 \centering
\subfloat[Tracking in the stable configuration with different noise amplitude] {
\includegraphics[width=0.9\linewidth]{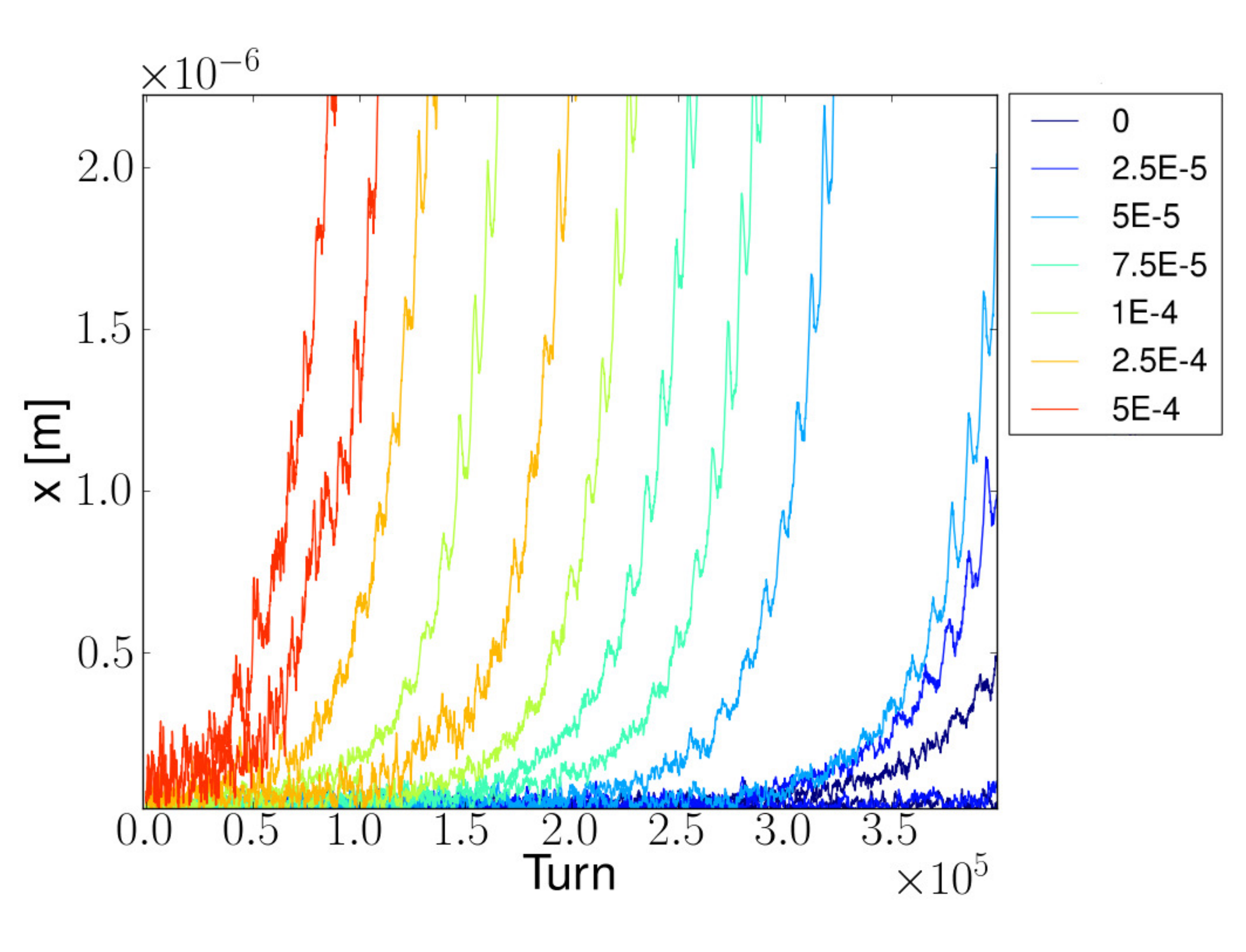}
}
\qquad
\subfloat[Latency time with a noise amplitude of \mbox{$5\cdot10^{-5}~\sigma_{x'}$}] {
\includegraphics[width=0.8\linewidth]{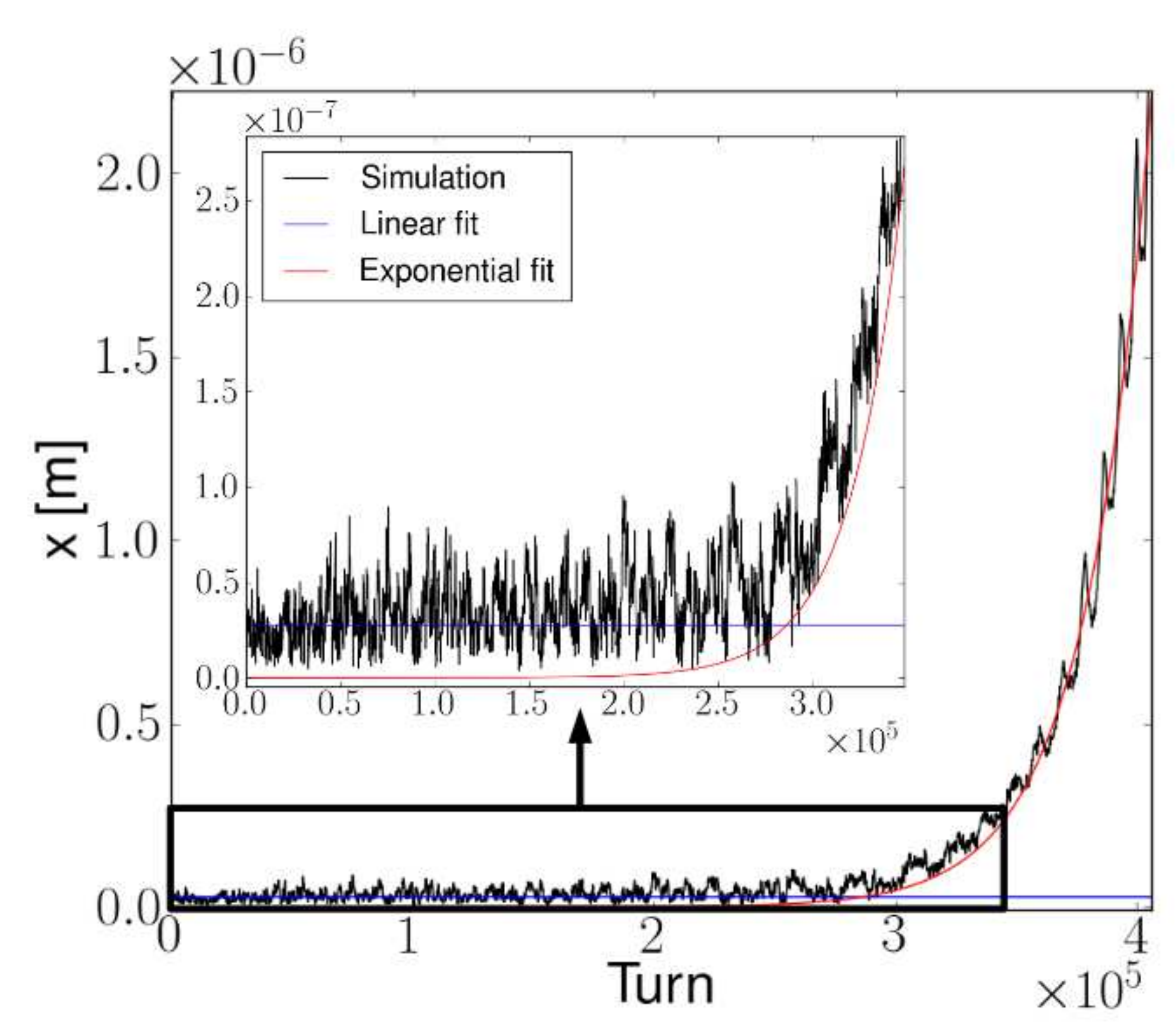}
}
\caption{Tracking simulations in the stable configuration (Fig.~\ref{fig-study case}) rendered unstable by wideband noise. The linear fit is done on the first part, and the exponential fit during the instability, which indicates a latency of $2.8\cdot10^5$ turns before an instability of rise time $2.7\cdot10^4$ turns.}
\label{fig-study case noise}
\end{figure}

The effect illustrated by this simple model can be simulated in more realistic scenarios. In particular, the COMBI code~\cite{COMBI}, extended with an impedance model, was used to model the action of external noise on a beam stabilized by both amplitude detuning and a transverse feedback. We consider a study case, where a single bunch is rendered unstable by negative chromaticity. This study case intends to mimic a situation similar to the LHC at the end of the squeeze where the beams have to be stabilized by high damper gain and high octupole current, while relaxing significantly the need for computational power due to the large number of bunches in this case. The results are therefore not meant to be compared to measurements, but rather to illustrate an effect that should be studied in more realistic cases in the future. The upper dot in Fig.~\ref{fig-study case}\subref{fig-study case sdiag} shows the coherent tune shift of the most unstable mode, with a large positive imaginary part. The imaginary part is brought down to the lower dot using a transverse feedback. The line represents the stability diagram provided by an octupolar amplitude detuning. In this configuration, this analysis suggests that both the damper and the octupoles are required to stabilize the beam, which is confirmed by tracking (Fig.~\ref{fig-study case}\subref{fig-study case rise}), showing the oscillation amplitude as a function of time for the different scenarios. The stable configuration can, however, be rendered unstable by a wideband noise, i.e. a random kick is applied every turn, the kick being constant over the bunch length. Indeed, the response to the wideband noise is more important at the mode frequencies as it cannot be entirely damped by a transverse feedback with finite gain. Wideband noise in this configuration has a similar effect on the diffusion as coloured noise, and therefore can compromise the stability from Landau damping, as described above. In particular, in Fig.~\ref{fig-study case noise}, the simulation of the stable case for different noise amplitude shows that the latency time before the instability takes off depends on the noise amplitude.

The extension of this study case to more realistic configurations including multibunch impedance and beam--beam interactions promises great challenges from the computational point of view, but is however not out of reach with current resources. In particular, one could expect that other sources of diffusion may counterbalance the effect described. Nevertheless, this analysis suggests that the external noise could play a critical role in configurations where both a transverse feedback and amplitude detuning are required to stabilize the beams.
\section{Conclusion}
Single bunch stability of beams colliding in arbitrary complex configurations can be investigated by numerically solving the dispersion integral, using amplitude detuning derived from single particle tracking simulations. A new code, based on MAD-X output, has been developed to assess the stability of the LHC beams in any operational configuration. This approach is however not suited to studying the stability of multibunch coherent modes.

The effect of distributions perturbed by external noise on the stability diagram has been investigated using the code mentioned above, and revealed very strong effects for modifications of the distributions well below the sensitivity of current measurements. It has been shown using multiparticle tracking codes that, indeed, a beam well stabilized by both amplitude detuning and transverse damper could be rendered unstable by the introduction of wideband noise.
\section{Acknowledgments}
The authors would like to acknowledge S. White and E. M\'etral for fruitful discussions.


\begin{thebibliography}{99}   
\bibitem{general-sdiag} A.W. Chao, \textit{Physics of Collective Beams Instabilities in High Energy Accelerators}, (John Wiley and Sons, New York, 1993).
\bibitem{non-gaussian-sdiag} E. M\'etral $et~al$., ``Stability Diagram for Landau Damping with a Beam Collimated at an Arbitrary Number of Sigmas,'' CERN-AB-2004-019-ABP (2004).
\bibitem{resonant-diffusion} A. Bazzani $et~al$., ``Diffusion in Hamiltonian Systems Driven by Harmonic Noise,'' J. Phys. A: Math. Gen. 31 (1998).
\bibitem{multibunch-mode} J.P. Koutchouk and F. Ruggiero, ``A Summary on Landau Octupoles for the LHC,'' LHC Project Note 163 (1998).
\bibitem{BBZ} S. White $et~al$., ``Beam--Beam and Impedance,'' these proceedings.
\bibitem{stability-BB-mode} Y.Alexahin, ``On the Landau Damping and Decoherence of Transverse Dipole Oscillations in Colliding Beams,'' CERN SL-96-064 AP (1996).
\bibitem{madx} http://mad.web.cern.ch/mad/.
\bibitem{madx-footprint} W. Herr, ``Particle Tracking With MAD-X Including LHC Beam--Beam Interactions,'' LHC Project Note 344 (2004).
\bibitem{madx-footprint2} W. Herr, ``Tune Shift and Spread Due to Short and Long Range Beam--Beam Interactions in the LHC,'' CERN SL/90-06 (AP), LHC Note No. 119 (1990).
\bibitem{LHC-cases} X. Buffat $et~al$., ``Operational Considerations on the Stability of Colliding Beams,'' these proceedings.
\bibitem{distribution} A. Bazzani $et~al$., ``Effect of Colored Noise on the Betatronic Motion: A Possible Mechanism for Slow Extraction,'' Nonlinear and Stochastic Beam Dynamics in Accelerators - A Challenge to Theoretical and Computational Physics, L\"uneburg (1997).
\bibitem{COMBI} T. Pieloni, ``A Study of Beam--Beam Effects in Hadron Colliders with a Large Number of Bunches,'' EPFL Th\`ese No 4211 (2008).
\end{thebibliography}
\end{document}